\documentclass[twocolumn,prb,showpacs,aps,epsf]{revtex4}
\begin{document}
\title{Raman spectra in epitaxial thin films $La_{1-x}Ca_xMnO_3$ (x = 0.33, 0.5) grown on different substrates}
\author{Y. M. Xiong$^a$, T. Chen$^a$, G. Y. Wang$^a$, X. H. Chen$^{a,b}$\footnote{Corresponding author, Electronic address: chenxh@ustc.edu.cn}}
\author{X. Chen$^b$ and C. L. Chen$^b$\footnote{Corresponding author, Electronic address: clchen@uh.edu}}
\affiliation{a) Structure Research Laboratory and Department of
Physics, University of Science and Technology of China, Hefei,
Anhui, 230026, P. R. China\\
b) Texas Center for Superconductivity and Advanced Materials and
Department of Physics, University of Houston, Houston, Texas
77204\\ }
\date{\today}

\begin{abstract}
Raman spectra of LCMO films grown on LAO (001), STO (001), and MgO
(001) substrates were studied at different temperatures. The
effect of temperature, doping level and strain on Raman spectra
are discussed in detail. With decreasing temperature, the changes
of Raman spectra are correlated with the disorder-order
transition, shuch as: paramagnetism to ferromagnetism, and
charge-ordering. The strain induced by lattice-substrate mismatch
affects the Raman spectra strongly. The mode induced by disorder
of oxygen defects is apparently observed in
$La_{0.67}Ca_{0.33}MnO_3$ film on STO due to the larger tensile
strain. While this mode can not be seen in the Raman spectra for
the films on LAO and MgO. A strong unsigned mode at about 690
$cm^{-1}$ is observed in all films except for
$La_{0.67}Ca_{0.33}MnO_3$ film on LAO, which is not observed in
bulk sample easily. It suggests that the mode is closely related
to the strain.
\end{abstract}

\maketitle

\section{Introduction}

Recently, $R_{1-x}A_xMnO_3$ ( R = rare earth, A = Ca, Sr, or Ba)
compounds have attracted much attention because of their colossal
magnetoresistance (CMR)\cite{Loudon,Aken,So,Biswas}. Particularly,
$La_{1-x}Ca_xMnO_3$ has a very rich phase diagram of ground states
with competing order parameters\cite{Jin,Corey}. The parent
compound is $LaMnO_3$, which belongs to the family of rotationally
distorted perovskites\cite{Iliev1} and is an anti-ferromagnetic
(AFM) insulator. With the increase of doped $Ca^{2+}$ content, CMR
has been observed in the Ca concentration of 0.2-0.5\cite{Tokura}.
In this doping range, the compound transfers from a
paramagnetic-insulator (PI) to a ferromagnetic-metal with
decreasing temperature. For the high doping range with 0.5 $\leq$
x $\leq$ 0.875, the doped charge carriers are localized and
ordered with stripe modulations at low
temperature\cite{Tomioka,Ramirez,Chen}. The compound at the phase
boundary with x = 0.5 undergoes first a FM transition, and a
simultaneous AFM and charge ordering (CO) transition at low
temperature. Thus, it is representative and important to
investigate the properties of compounds which are optimal doping
for Curie temperature ($T_c$) (x = 0.33) and the phase boundary
doping (x = 0.5).

A competition between Super-Exchange (SE)\cite{Goodenough} and
Double-Exchange (DE)\cite{Zener} interactions has long been used
to explain the resulting spin alignment and transport properties
of these manganese oxides. In addition, the strong electron-phonon
coupling was found to play an important role in the mechanism of
CMR\cite{Millis1,Millis2}. In particular, the Jahn-Teller (JT)
effect affects both the magnetic and transport properties in
${La_{1-x}Ca_xMnO_3}$ (LCMO) system\cite{Millis1,Millis2,Millis3}.
There are two types of distortions of the ideal perovskite
structure in bulk samples. One is the rotational-like distortions
of the $MnO_6$ octahedra due to the mismatch between the averaged
ionic radii $r_A$ of the A-site species and the ionic radius
$r_{Mn}$ of Mn. It is one of the strain effects induced by
chemical pressure\cite{Steenbeck,Ibarra}. The other is JT
distortions of $Mn^{3+}O_6$ octahedra. It is a strong
electron-phonon coupling. The strain can also be caused by the
lattice-substrate mismatch\cite{Biswas,Campillo}. This kind of
epitaxial strain is different from the strain induced by
hydrostatic\cite{Teresa} or chemical
pressure\cite{Steenbeck,Ibarra}, since an in-plane strain
generally accompanies an out-of-plane strain with different sign,
which can cause new electronic behavior did not found in bulk
materials of the same chemical composition\cite{Biswas}. The
interplay between substrate and film allows the modification of
properties and can even enhance the magnetoresistive
effect\cite{Jin,Helmolt}. This suggested a possible device
applications with these films based on the sensitivity to magnetic
fields\cite{Prellier}. There are two mechanisms by which
substrates modify film properties. The first one is associated
with the lattice distortion of epitaxially grown films. The
substrate induces both bulk and biaxial strains in the film, which
alters the physical properties of
films\cite{Wang1,Wang2,Li,Miniotas}. The other one is associated
with the dynamics of film growth and the manner of strain
release\cite{Campillo}, which induce phase separation and
inhomogeneities in films\cite{Lu,Biswas2}. These suggest that the
interesting phenomena maybe observed in highly strained thin films
of manganite. So we grew films on substrates with different
lattice parameters, and investigated the change of their
properties to study the effect of strain induced by substrate.

Raman scattering was proved to be a useful tool for the study of
various vibration and electronic excitations, phase transitions,
lattice distortions, etc.. Thus, this technique affords a means by
which energy and symmetry information regarding phonon, electron,
and spin excitations can be simultaneously explored. It is well
known that there are no Raman-active of lattice vibrations in the
ideal cubic perovskite, all phonon modes originate from lattice
distortions. Therefore, we can use Raman spectroscopy to study the
lattice distortions in LCMO films which induced by the change of
temperature and the strain comes from lattice-substrate mismatch.
In addition, there are seldom work has been done on Raman spectra
of the thin films with composition of 0.5.

In order to study the effect of strain systemically, LCMO films
were deposited on (001)$LaAlO_3$ (LAO), (001)$SrTiO_3$ (STO), and
(001)MgO (MgO) substrates, respectively. The strain in each film
is different because the lattice parameters of these substrates
are different. We investigated the Raman scattering in these films
at different temperatures to study the change of phonon vibration
induced by temperature, lattice-substrate mismatch and Ca
concentration.

\section{Experiments}

Thin films of $La_{1-x}Ca_{0.33}MnO_3$ (x = 0.33 and 0.5), about
200-300 nm in thickness, were grown on $LaAlO_3$ (001), $SrTiO_3$
(001), and MgO (001) substrates by pulsed laser deposition. The
films were deposited at 800 $^oC$ in 400 mTorr oxygen partial
pressure without subsequent post-annealing. The x-ray diffraction
(XRD) measurements were carried out in Rigaku D/max-$\gamma$A
x-ray diffractometer with graphite monochromatized Cu $K_\alpha$
radiation ($\lambda$=1.5406\AA). Raman spectra were obtained on a
LABRAM-HR Confocal Laser MicroRaman Spectrometer using the 514.5
nm line from an argon-ion laser. The data were fitted with
Lorentzian function. The magnetic susceptibility measurement was
carried out with a superconducting quantum interference device
magnetometer(MPMS-5). The high pressure oxygen anneal were carried
out in High Pressure Oxygen Furnace (Morris Research) at 195 atm
and 500 $^oC$ for 24 hours.

\section{Results and Discussion}

Figure 1 shows XRD pattern for films grown on three substrates
with x = 0.33 and 0.5. We labelled $La_{0.67}Ca_{0.33}MnO_3$ as
LCMO1 and $La_{0.5}Ca_{0.5}MnO_3$ as LCMO2, respectively. The
films were of single phase and highly crystallized as seen from
the sharp and intense diffraction peaks. Generally, at the best
growth conditions, the single-crystalline can be obtained in
orthorhombic structure with the space group of Pnma due to the
tilting of $MnO_6$ octahedra. In the films, the orientation of the
orthorhombic axes in respect of the interface depends on the type
of the substrate\cite{Lu,Tatsi}. It has been found that LCMO1
films were [110] oriented on LAO substrate and mixed [001] and
[110] oriented on STO substrate\cite{Lu}. Because the splitting
between the (110) and (002) reflection of orthorhombic LCMO is too
small to be resolvable (see JCPD No. 49-0416), the peaks of the
films are indexed based on the single perovskite unit cell.
According to the XRD data, we calculated the out-of-plane lattice
parameters of films and substrates and listed them in the Table I.
These results are consistent with the data reported in previous
references\cite{Wang1,Schmahl,Mathur,Abrashev}. As shown in Table
I, the lattice parameter of $LaAlO_3$ is 3.7916 $\AA$, smaller
than the lattice parameters of bulk crystals value of 3.86 $\AA$
for $La_{0.67}Ca_{0.33}MnO_3$ (LCMO1)\cite{Mathur,Bibes} and 3.84
$\AA$ for $La_{0.5}Ca_{0.5}MnO_3$ (LCMO2)\cite{Abrashev}. Thus,
the lattices of films on LAO undergo a simultaneous in-plane
compression and an out-of-plane elongation due to the in-plane
compressive strain induced by LAO substrate. In contrast, lattices
of films grown on STO should undergo an in-plane tensile strain
accompanying by a shrinkage along the out-of-plane orientation
because its lattice parameter (3.9026 \AA) is larger than the bulk
value. It has been reported that the lattice-substrate mismatch
affects the Mn-O bond length slightly in epitaxial LCMO films
grown on different substrate, the change of Mn-Mn interatomic
distance is controlled by altering the Mn-O-Mn
angle\cite{Miniotas}. Thus, the in-plane compression and
out-of-plane elongation of films on LAO are mainly achieved by the
out-of-phase rotation of $MnO_6$ octahedra. The same is true of
the films on STO, the lattice distortion of films on STO is also
achieved by the out-of-plane rotation of $MnO_6$ octahedra.

It is expected that the out-of-plane lattice parameters of LCMO
films grown on MgO should be smaller than that of films grown on
STO, because the lattice parameter of MgO is larger than that of
STO. While, it is interesting to note that the lattice parameters
for the films grown on MgO are very close to the values from the
LCMO bulk materials(see Table I). The Raman spectra also indicate
that no strong strain effect appears in the MgO samples. These
phenomena probably result from the strain-relieving
mosaics\cite{Wollschlager} that has recently been discovered from
the $CeO_2$ thin films on (001) MgO\cite{Chen1}. Theoretically, if
the lattice mismatch is sufficiently large, a coincidence lattice
can be formed to accommodate the large strain through slightly
rotating film relative to the substrate. This rotational mismatch
allows the film to assume its bulk lattice parameter and structure
for resulting in a quasi-epitaxial behavior. Details about the
mechanisms are under studied and will report later on. It is
noticed that the out-of-plane lattice parameter increases when Ca
concentration changes from 0.33 to 0.5 in the films grown on LAO.
While in those grown on STO and MgO, their out-of-plane lattice
parameters decrease. It may be induced by the different strain in
each sample. The discussion on it will be carried out in detail
later. In the following, we will discuss the effects of
temperature, lattice-substrate mismatch and doping on Raman
spectra.

Figure 2 and 3 show the Raman spectra of LCMO films with Ca
concentration of 0.33 and 0.5 on three substrates at the
temperatures of 293, 260, 240, 220, 200, 180, 160, 140, and 100 K,
respectively. It is shown that the Raman modes are mainly located
in three intervals at about 220$\sim$250 $cm^{-1}$ ($\omega_1$),
450$\sim$520 $cm^{-1}$ ($\omega_2$), and 610$\sim$720 $cm^{-1}$
(the lower frequency mode is labelled as $\omega_3$ and the higher
frequency mode is labelled as $\omega_4$). $\omega_1$ belongs to
$A_g(2)$ mode (rotational mode), and is proportional to the angle
of long-rang coherent static tilt of $MnO_6$ octahedra. The higher
frequency modes $\omega_2$ and $\omega_3$ are assigned to
JT-active vibrations. $\omega_2$ belongs to $A_g(3)$ mode (bending
mode) which is determined by the mean distance between La/Ca and
apical oxygens. According to previous reports, $\omega_3$ with
$B_{2g}(1)$ symmetry is an in-phase oxygen
stretching\cite{Podobedov,Carron,Abrashev1}. Podobedov et
al\cite{Podobedov}. considered this band is a disorder-induced
Raman feature and expect its origin to reside in a lattice defect
due to an existing oxygen deficit. This conclusion is supported by
the less intensity of about 600 $cm^{-1}$ band in annealed
samples\cite{Podobedov}. Malde et al.\cite{Malde} observed that
the frequency of this mode shifts with the variation of annealing
oxygen partial pressure. This seems to support the origin of this
mode from the oxygen deficit. The relative intensity of this
disorder-induced Raman band may serve as a good indication of the
oxygen deficit in LCMO\cite{Podobedov}. In other
articles\cite{Abrashev1,Abrashev3,Liarokapis,Dediu}, these authors
believe this band from the JT distortion of $MnO_6$ octahedra.
Therefore, it is necessary to study the origin of the 600
$cm^{-1}$ mode.

According to previous report in $La_{0.67}Ca_{0.33}MnO_3$ film on
LAO\cite{Podobedov}, a higher frequency mode at about 660
$cm^{-1}$ comes from the degradation of the sample surface area.
While the strong peaks at about 690 $cm^{-1}$ and the weak band at
about 610 $cm^{-1}$ in our samples are inconsistent to the results
reported by Podobedov et al\cite{Podobedov}. In order to confirm
the origin of the two peaks, we annealed our samples in high
pressure oxygen. Because the two peaks are weak in Raman spectra
of LCMO1/LAO, it is hard to distinguish the annealing effect. We
compared the Raman spectra of LCMO1 films on STO and MgO, and
LCMO2 on LAO. Figure 4 shows the Raman spectra of as-grown and
annealed LCMO1 films on STO and MgO, and LCMO2 film on LAO,
respectively. These samples were annealed in high pressure oxygen
at 195 atm and 500 $^oC$ for 24 hours. As shown in Fig. 4, for the
intensity of $\omega_4$, no obvious weakening can be observed
after annealing, even it increases in LCMO1/MgO. Thus the
potential origin of $\omega_4$ from oxygen deficit could be
incorrect. $\omega_4$ is unsigned and unfamiliar in Raman spectra
of LCMO samples. Detailed discussions are carried out later. In
our samples, the change of $\omega_3$ is hardly distinguished
because of the weaker relative intensity than $\omega_4$. In the
bottom panel of Fig.4, there is a weak peak at about 610 $cm^{-1}$
in as-grown LCMO2/LAO. After annealing in high pressure oxygen,
this peak disappears. It may indicate a decrease of oxygen
deficit. Therefore, our results seem to support the origin of 600
$cm^{-1}$ from the oxygen deficit. However, no apparent change for
$\omega_3$ is observed in LCMO1/STO after annealing in high
pressure oxygen. It suggests that the oxygen deficit in LCMO1/STO
is very stable due to a large tensile strain. Thus, we believe the
mode at about 600 $cm^{-1}$ is correlated with the oxygen deficit
and affected by the strain strongly.

\subsection{Effect of Temperature}

In the Fig.2, the Raman spectra of $La_{0.67}Ca_{0.33}MnO_3$ films
were measured at the temperatures of 293, 260, 240, 220, 200, and
100 K. In Raman spectra of LCMO1/LAO film, a much sharp peak
appears at about 486 $cm^{-1}$, and this feature is almost
unchanged with changing temperature. Compared with the Raman
spectra of LAO substrate\cite{Abrashev1}, this sharp peak is
attributed to LAO substrate, identified by asterisks($*$). The
Raman spectra feature from film near 480 $cm^{-1}$ is broad and
hardly distinguished because of the overlap of peaks from film and
LAO substrate. The high frequency modes $\omega_3$ and $\omega_4$
of film grown on LAO are weak, and temperature dependence of
frequency shift can not be observed clearly. In bulk samples with
Ca concentration in the region of 0.3 $<$ x $<$ 0.4, the
$\omega_3$ mode is weak. The weakening of this mode can be
correlated with the reduction of JT distortion, introduced by the
presence of Ca\cite{Liarokapis}. In the film grown on LAO, Raman
spectra are similar to those in bulk samples. The high frequency
modes are weak due to the present of $Mn^{+4}$ ions which is JT
inactive. Figure 5 shows temperature dependence of the frequency
shift of Raman modes for LCMO1 films. In the Fig. 5a, the
$\omega_1$ mode for LCMO1/LAO hardens with decreasing temperature.
In earlier reports, there is one mode in this frequency range for
LCMO samples\cite{Abrashev1,Iliev2,Amelitchev}, and it is with
$A_{g}(2)$ symmetry and attributed to the rotation of $MnO_6$
octahedra. In the film on LAO, the $MnO_6$ octahedra rotate to
match the distortion of lattice induced by lattice-substrate
mismatch. At room temperature, the in-plane shrink and
out-of-plane elongation of lattices is achieved by the rotation of
$MnO_6$ octahedra. With decreasing temperature, the hardening of
$\omega_1$, indicating the increase of tilt angle of $MnO_6$
octahedra, is maybe caused by the increase of lattice-substrate
mismatch. The peak of $\omega_1$ sharpens slightly below 220 K. As
shown in the inset of Fig.2a, the magnetic phase transition can be
observed clearly, the Curie Temperature ($T_C$) is about 220 K.
The sharpening of $\omega_1$ peak has been reported earlier by
Podobedov et al\cite{Podobedov} and attributed to the spin-lattice
coupling in the spin-ordered FM state. Recently, another
explanation of this narrowing of $\omega_1$, could be the increase
of long range structural coherence due to the reduction of the
dynamic JT distortion\cite{Abrashev1,Irwin}.

According to the previous reports, the Raman feature $\omega_2$
could contain several peaks with different
symmetry\cite{Carron,Iliev2,Iliev3}, its temperature dependence is
complex due to the overlap of these modes. The different
temperature behaviors of these modes induce a shift of the broad
$\omega_2$ band. As shown in Fig. 2a, with decreasing temperature
some much sharp peaks appear and their intensities become more
strong at the temperatures below $T_C$. This is consistent with
that reported in $La_{0.7}Ca_{0.3}MnO_3$/LAO film\cite{Abrashev1}.
The 441.4 $cm^{-1}$ line is classified to the bending $E_g$ mode
for rhombohedral $La_{0.93}Mn_{0.98}O_3$\cite{Iliev2}. These sharp
peaks may be correlated with the electron or spin excitations
existing in the metallic-state phase. Recently, another
explanation was reported that these much sharper Raman lines
corresponding to the $\Gamma$-point Raman phonon of the ordered
structure reflects the disorder-order transition, concomitant with
the PM to FM and insulator-metal transitions\cite{Iliev3}.
However, no sharp peaks appear when the sample enters FM state in
the film on STO. In addition, because the $E_g$ modes always
appear in rhombohedral symmetric structure\cite{Iliev2,Abrashev2},
the appearance of these sharp peaks maybe indicate a structure
transition from orthorhombic lattice to rhombohedron. These sharp
peaks in the 400 to 450 $cm^{-1}$ range at low temperatures also
appears in bulk samples with composition of about
0.33\cite{Liarokapis}. This indicates that a compressive strain
can not affect the evolvement of lattice with decreasing
temperature. While a tensile strain affects the lattice strongly.
These sharp peaks in the 400 to 500 $cm^{-1}$ range always appear
together in the Raman spectra of
LCMO\cite{Abrashev1,Liarokapis,Iliev2,Iliev3,Irwin}. Based on
their similar behavior with the temperature and substrates in our
case and the earlier reports in films and bulk
samples\cite{Abrashev1,Liarokapis}, we attribute these narrow
peaks to the same symmetry.

The LCMO1 film grown on STO undergoes a tensile strain, whose
Raman spectra are shown in the Fig. 2b. The in-plane lattice
expands due to the tensile strain, at the same time out-of-plane
lattice shrinks. This kind of anisotropic distortion of lattices
should induce the tilt of $MnO_6$ octahedra. The frequency of
$\omega_1$ mode decreases with decreasing temperature which may be
induced by decreasing tilt angle of $MnO_6$ octahedra due to the
reduction of lattice-substrate mismatch (see Fig. 5a). The
intensity of $\omega_1$ increases apparently above 240 K, and
decreases dramatically with the decrease of temperature below 240
K. In contrast, the full width at half maximum (FWHM) decreases
firstly above 240 K, then increases with decreasing of
temperature. The strongest intensity of $\omega_1$ mode at 240 K
indicates a strongest static JT distortion at this temperature.
The dramatic decrease of intensity below 240 K should be induced
by spin-lattice coupling in spin-ordered FM state. As shown in
Raman spectra of LCMO1 film grown on STO, $\omega_2$ is absent at
room temperature, then appears and becomes more obvious at low
temperature. These changes are caused by the increase of $MnO_6$
octahedra bending at the low temperatures. The frequencies of
$\omega_2$ and $\omega_3$, which are correlated with JT
distortion, shift up as temperature decreases. This is induced by
lattice shrink due to the decrease of temperature. As shown in
Table I, the out-of-plane lattice parameter of film grown on STO
is the least, which indicates the stretching of in-plane lattices.
Therefore, the mean distance between La/Ca and apical oxygens is
larger than that in bulk sample. With decreasing temperature, the
distance between La/Ca and apical oxygens decreases continually,
which leads to an hardening of the frequency in $\omega_2$. As
shown in the Fig.2b, the intensities of $\omega_2$ and $\omega_3$,
increase with decreasing temperature compared with $\omega_1$. It
is due to the change from small-polaron-dominated transport in the
paramagnetic (PM) phase to large-polaron (metallic) transport in
the FM phase\cite{Yoon}. No obvious change in $\omega_4$ is
observed when the sample enters FM state. It suggests that
$\omega_4$ is independent of the PM to FM transition.

In the spectra of LCMO1/MgO, the frequency of $\omega_1$ hardens
slightly with lowering temperature. According to the XRD pattern
(see Fig. 1), the lattice-substrate mismatch is slight in film on
MgO. In addition, the intensity of $\omega_1$ in LCMO1/MgO
increases above 220 K, reaches the maximum at 220 K, then
decreases dramatically. This behavior is similar to the case of
LCMO1/STO and may be correlated with the PM to FM transition, as
observed in Raman spectra of LAMO1/STO. $\omega_2$ feature hardens
with decreasing of temperature above 220 K, then softens again
(see Fig. 2c). These phenomena may be due to the coexistence of
several modes in this interval. These modes have different
temperature behaviors because of their different symmetry. As
previously reported in $RMnO_3$ (R = Ca, Pr, Nd, Tb, Ho, Er and Y)
samples, this feature contains the bending and antisymmetric
stretching modes\cite{Carron}. The frequency of antisymmetric (AS)
stretching mode simply depends on the Mn-O bond length. The
frequency of bending mode strongly depends on the mean value of
the shortest distance between La/Ca ion and apical oxygen. With
decreasing temperature, the Mn-O bond length shrinks slightly,
while the La/Ca-O distances change obviously. Thus, these
different temperature effect of each mode induce a complex
behavior of $\omega_2$ feature. As shown in Fig. 2c, $\omega_3$
band can not be observed, which indicates a little oxygen deficit
in this film.

In the Fig. 3, the Raman spectra of $La_{0.5}Ca_{0.5}MnO_3$ films
are shown at the temperatures of 293, 250, 200, 180, 140, 100, and
83 K, respectively. The bulk LCMO sample with Ca concentration of
0.5 is at the phase boundary between FM and charge ordered AF
state. According to the previous report\cite{Iliev2}, with
decreasing temperature a competition between the ferromagnetic and
antiferromagnetic ordering occur. In a narrow region
(0.44$<$x$<$0.54) upon cooling, the compound becomes first
ferromagnetic, then antiferromagnetic, in particular, for x = 0.5,
$T_C$ = 225 $\sim$ 265 K, $T_N$= 130 $\sim$ 160 K. Below $T_N$
orbital and charge ordering takes place and a superstructure of
$P2_1$/m symmetry is formed supported by the much richer Raman
spectra feature in this sample than that of orthorhombic
$LaMnO_3$\cite{Radaelli2}. In the Raman spectra of LCMO2 film on
LAO, the peak at about 486 $cm^{-1}$ is from LAO substrate,
identified by asterisks. The frequency of $\omega_1$ hardens with
decreasing temperature (see Fig. 6) due to the increase in tilt
angle of $MnO_6$ octahedra, as observed in LCMO1 film. At the same
time, the intensity of $\omega_1$ increases slightly above 180 K,
then decreases continuously with decreasing temperature. At low
temperature, several additional peaks present at 425, 446, and 469
$cm^{-1}$ in LCMO2 on LAO. Similar spectral shape at low
temperature has been reported in bulk $La_{0.5}Ca_{0.5}MnO_3$ by
Iliev et al.\cite{Iliev2} and Abrashev et al.\cite{Abrashev}. It
is believed that the presence of these sharp peaks is induced by
the lowering symmetry as samples enter CO state. The peak of
$\omega_3$ mode is weak and can not be observed clearly. This
weakening of disorder-induced band does not indicate a less oxygen
deficit in LCMO2. It has been reported in bulk samples that the
distorted volume reaches its maximum for disordered
$Mn^{3+}$/$Mn^{4+}$ composition with $\sim$ 30 $\%$
$Mn^{4+}$.\cite{Iliev3} The decrease in intensity of the
disorder-induced bands at higher doping and the appearance of some
sharp Raman features can be explained by increasing the volume of
domains with short-range charge/orbital-ordered\cite{Abrashev}. As
reported in bulk sample of $La_{0.5}Ca_{0.5}MnO_3$, Raman
structures are weak at room temperature, the structure at $\sim$
480 and $\sim$ 600 $cm^{-1}$ develop and become strong with
decreasing temperature below CO
temperature.\cite{Iliev2,Abrashev,Granado} Compared with the bulk
sample, the Raman spectra is similar to that for bulk sample
except that obvious peaks appear at $\sim$ 235 and $\sim$ 690
$cm^{-1}$ at room temperature. With decreasing temperature, the
intensity of the mode at at $\sim$ 480 $cm^{-1}$ increases and the
Raman shift shows a systematical change. However, no apparent peak
at $\sim$ 600 $cm^{-1}$ is observed with cooling the temperature
to 83 K. The different spectra from bulk sample could arise from
the strain induced by the lattice mismatch between the film and
substrates. In addition, the peak of $\omega_4$ enhances and
hardens gradually with decreasing temperature and no obvious
change is seen when the sample enters charge ordering state.

The temperature dependence of Raman spectra for LCMO2/STO are
shown in Fig. 3b. The intensity of $\omega_1$ increases above 160
K, then decreases dramatically below 160 K as temperature
decreases. At the same time, the frequency of this mode softens
first slightly with decreasing temperature, then hardens below 160
K (see Fig. 6). The decrease in intensity of $\omega_1$ below 160
K is contrary to the previous report in bulk
samples\cite{Abrashev,Granado}. In bulk samples, the modes enhance
as the system enters CO state due to the lowering symmetry.
Thereby, the anomalous change in the intensity of $\omega_1$
should be induced by the strain. In additional, the anomalous
change only occurs in the band of $\omega_1$. It suggests that the
strain effect affects the tilt angle of $MnO_6$ octahedra
strongly. $\omega_2$ becomes obvious and narrow gradually with
decreasing of temperature. The slight hardening of $\omega_2$
implies that the distance between La/Ca and apical oxygens shrinks
slightly. The frequencies of high frequency modes increase with
decreasing temperature due to the compression of lattices. The
mean distance between La/Ca and apical oxygens and the length of
Mn-O bond are elongated fully in the plane parallel to the
substrate. Therefore, the lattices shrink continually as
temperature decreases, then all high frequency modes shift up.

In LCMO2/MgO sample, $\omega_1$ is broad and weak, and weakens
with lowering temperature. It implies that there are a small
quantity of tilted $MnO_6$ octahedra in this sample, which become
much less with decreasing temperature. According to the XRD
pattern, the lattice distortion of film on MgO is the least. The
lack of tilted $MnO_6$ is consistent with this result. The change
of $\omega_2$ is similar to that in LCMO1 film on MgO and the
temperature corresponding to the strongest intensity is 140 K. The
reason is the same as that in the case of x = 0.33. The intensity
of $\omega_4$ increases continuously to the maximum, and FWHM
narrows to the minimum at 100 K. Then the intensity decreases
dramatically.

In summary, we discussed the variation of Raman spectra of LCMO
films with changing temperature in this part. Although, they were
deposited on the different substrates, their Raman spectra are all
correlated with the transition of magnetic phase from disorder to
order. These transitions usually are accompanied by the change of
Jahn-Teller distortion. With decreasing temperature, the Raman
spectra change obviously at the phase transition temperature.
$\omega_1$ mode, which is correlated with the tilt angle of
$MnO_6$ octahedra, hardens or softens with lowering temperature
due to the change of lattice-substrate mismatch. As samples enter
the ordered FM state, the peak of $\omega_1$ band sharpens in
LCMO1/LAO due to the increase of spin-lattice coupling. In LCMO1
on STO and MgO, the intensity of $\omega_1$ decreases dramatically
because of the spin-lattice coupling too. $\omega_2$ and
$\omega_3$ are correlated with JT distortion. The behavior of
$\omega_2$ feature is complex due to the coexistence of several
modes. These modes with different symmetry have different behavior
with lowering temperature. As previous report\cite{Carron}, in
this interval, the bending mode would change obviously, while the
stretching mode would change slightly as temperature decreases.
$\omega_3$ is a indicator of oxygen deficit in LCMO1 films and
weakens in LCMO2 films due to the increasing the volume of domains
with short-range charge/orbital-ordered\cite{Abrashev}. $\omega_4$
does not arise from oxygen deficit in films. Detailed discussions
about this mode are carried out later.

\subsection{Effect of the Lattice-Substrate Mismatch}

To study the effect of strain in these films, we compare the Raman
spectra of films on different substrates. As shown in Fig. 2,
there are four main peaks centered at the frequency about 228,
450, 625 and 675 $cm^{-1}$ in all films. At room temperature, the
frequency of $\omega_1$ in the films on MgO is less than that of
STO, and that of LAO is least (see fig. 5). With the change of the
mismatch between film and substrate, the lattices of film undergo
an in-plane elongation (compression) and a simultaneous
out-of-plane compression (elongation). In the film on STO, the
frequency of $\omega_1$ is the largest one due to the larger
tensile train in this film. With decreasing temperature,
$\omega_1$ softens in LCMO1/STO film, while it hardens in
LCMO1/LAO film. The thermal expansion coefficients of LCMO films
are not available. The thermal expansion coefficients of LAO
(1$\times$$10^{-5}$ $K^{-1}$), STO (0.9 $\times$$10^{-5}$
$K^{-1}$), and MgO (1.1 $\times$$10^{-5}$ $K^{-1}$) are close to
each other\cite{Wang1}, the thermal expansion contribution to the
lattice distortions should be similar for the films on three
substrates. To explain the contrary shift of $\omega_1$ band in
the films on different substrate, it is reasonable that we assume
the thermal expansion coefficients of substrates are larger than
that of the film. In LCMO1/LAO film, the lattice mismatch
increases with decreasing temperature due to the above assumption.
Thus, the tilt of $MnO_6$ octahedra is harder at low temperature.
In LCMO1/STO film, the mismatch decreases as temperature
decreases. $\omega_1$ hardens slightly in film on MgO with
decreasing temperature due to the slight strain in this film at
room temperature. Therefore, the tilt angle of $MnO_6$ octahedra
is partly affected by the strain in LCMO1 films. In LCMO1 film on
STO, no peak appears around 480 $cm^{-1}$ at room temperature. It
implies that there is no $MnO_6$ octahedra bending or the banding
mode is too weak to be observed. In addition, there are no
additional sharp peaks appear in the films on STO and MgO at low
temperature. While these sharp peaks appeared in the 400 to 450
$cm^{-1}$ range at low temperature are observed in film on LAO and
$La_{0.67}Ca_{0.33}MnO_3$ bulk samples\cite{Liarokapis}. It proves
that these much sharp peaks are determined by the lattice
distortion, which can be modified by the type of strain. According
to the above discussion, only in-plane compressive strain could
induce a structural transition from orthorhombic lattice to
rhombohedron. As shown in Fig. 2 and Fig. 3, both the $\omega_2$
band in films on MgO hardens above a certain temperature and
softens below this temperature, no similar behavior is observed in
films on LAO and STO. Thereby, it suggests that this complex
frequency change of $\omega_2$ depend on the type of substrates
strongly. In the region of $\omega_2$ band, the various modes
coexist. Each mode with different symmetry should be affected by
strain differently. In the films on LAO and STO, the effect of
strain induces the monotonic change of frequency. While in the
films on MgO, the strain effect is not dominant.

At high frequencies, the LCMO1/LAO sample has a broad peak at
about 670 $cm^{-1}$, which contains two peaks. The lower frequency
mode is correlated with the oxygen deficit in the films and may
serve as a good indicator of the oxygen deficit in
LCMO\cite{Podobedov}. The strong intensity of this mode in
LCMO1/STO film indicates that the oxygen deficit is the largest in
film with tensile strain. As shown in Fig. 4, in the bottom panel,
there is a weak peaks at about 610 $cm^{-1}$ for as-grown
LCMO2/LAO film. After annealing in high pressure oxygen, this peak
disappears. It may indicate a decrease of oxygen deficit induced
by the annealing process. In the top panel of Fig. 4, the
$\omega_3$ mode is unchanged after annealing. It suggests that the
lattice defects in LCMO1/STO film are intrinsic defects induced by
larger tensile strain, which is hardly reduced by the annealing
process. It is found that Raman spectra of our samples contain a
peak at about 690 $cm^{-1}$. This peak is observed in both LCMO1
and LCMO2 films. According to previous reports on bulk and film
LCMO samples, this high frequency mode is unassigned and may be
related to the Ca doping, which originates from
impurity\cite{Podobedov,Malde,Pantoja}. There are mainly two
models in previous reports. In the first model, because this peak
often presents in spectra of annealed $LaMnO_3$ samples and
disappears in those polished samples\cite{Podobedov2}. It is maybe
a sign of simple oxides like that of lanthanum or manganese
originated from long-term degradation of the sample surface. In
films on STO and MgO, this peak has a higher intensity, which
suggests a more strongly degradation of these film's surface.
However, the dependence of strain for this mode observed in our
samples is hardly to explain within the above model. The frequency
of this mode hardens when the sign of strain changes from tensile
to compressive. The ideal epitaxy is preserved only within about
the first 25 nm from the substrate-film interface\cite{Lu}. The
strain releases with increasing of thickness. If this mode is a
representation of simple oxides on the sample surface, the strain
effect should be small. The strong dependence of strain for this
mode suggests that this mode should not relate with the oxides on
the sample surface. Another reasonable explanation of this peak
could be the second-order Raman
scattering\cite{Podobedov2,Podobedov3}. The comparable intensity
of the second-order Raman spectra indicates both a relatively
small distortion of a nearly cubic lattice and a disorder present
in the perovskite structure. However, according to the XRD pattern
shown above, the lattice distortion of film on MgO is a least one.
In addition, it is hardly to be considered as a second-order due
to such a high relative intensity of this peak in film on STO and
MgO. According to the above discussion, we consider that this new
peak is correlated with the interaction between lattice and
substrate. This additional higher frequency mode is also observed
in $Pr_{1-x}Ca_xMnO_3$ samples, caused by the lowering of the
symmetry as a result of unit cell doubling along the $a$
axis\cite{Tatsi,Cox}. Therefore, one reasonable possibility of the
appearance of this mode comes from a lowering of the symmetry
induced by strain. In addition, it is noticeable that this mode
enhances and hardens in both LCMO1 and LCMO2 films with decreasing
temperature. It seems to be independent of the Curie temperature
($T_C$) or the charge ordering temperature ($T_{CO}$), and to be
decided by the strain simply. Therefore, we speculate that this
mode reflect the information of the layers near the
lattice-substrate interface. More detailed research about this
peak need more experiments.

Similarly, the Raman spectra of $La_{0.5}Ca_{0.5}MnO_3$ films are
also strongly depend on the type of strain. As shown in the Fig.3,
obvious differences can be observed between the films on different
substrates. The intensity of $\omega_1$ is weak in LCMO2/LAO film
with a compressive strain at room temperature, and is the
strongest in LCMO2/STO sample with a tensile strain, then is the
most weak in film on MgO. The frequency changes complexly with
decreasing of temperature (see Fig. 6). Its behavior depends on
the type of strain. The changes of $\omega_2$ in LCMO2 films are
similar to those of LCMO1, strongly depend on the type of strain.
Some sharp peaks appeared at low temperatures suggests a lowering
symmetry, as observed in $La_{0.5}Ca_{0.5}MnO_3$ bulk samples. It
seems to imply a transition of CO state. The frequency of
$\omega_4$ softens as the strain changes from a compressive one to
a tensile one. It is consistent with the change of spectra in film
with x = 0.33. It further proves that $\omega_4$ mode is
correlated with the lattice-substrate mismatch, namely strain in
films. In addition, $\omega_3$ can be observed clearer only in STO
sample due to the large tensile strain, which leads to more oxygen
deficit in the film during the growth process. These oxygen
deficits induce by larger tensile strain are intrinsical and
hardly suppressed.

There is a relaxation of the strains from the interface towards
the surface. And the Raman scattering probes mainly the surface.
It is probably that we observe different effect from strain by
varying the laser wavelength. Unfortunately, we can not change the
laser wavelength because of the limit of equipment. In our case,
the thicknesses of the films are same. Thus, our discussion did
not refer to the thickness of the films.

\subsection{Effect of Ca Concentration}
We compared the Raman spectra of samples with different Ca
contents (x = 0.33 and 0.5) on the same substrate at different
temperatures. It should be noticed that the out-of-plane lattice
parameter decreases as Ca concentration increases in films on STO
and MgO. While the lattice parameter increases in film on LAO.
This difference in the change of lattice is induced by the
different type strain in each film. In the film on LAO, the
increase of out-of-plane lattice parameter with the increase of Ca
concentration is induced by the following two reasons. One is the
increase of Mn-$O_{apical}$-Mn angle. Liarokapis et
al.\cite{Liarokapis} had reported that the Mn-$O_{apital}$-Mn
angle increases with the increase of Ca content in samples of
higher Ca concentrations. This point can be proved by the
softening of $\omega_1$ in LCMO2/LAO compared with LCMO1/LAO. The
other is the increase of $Mn-O_{apical}$ bond length. Radaelli et
al.\cite{Radaelli} had reported that the distance between Mn ions
and apical oxygens should increase with the decrease of A-site
ionic radius $<r_A>$ as $<r_A>$ is less than 1.24 $\AA$. The
$<r_A>$ for $La_{0.7}Ca_{0.3}MnO_3$ is 1.205 $\AA$. In contrast,
the out-of-plane lattice parameters of films grown on STO and MgO
decrease as Ca concentration increases. This decrease of
out-of-plane lattice parameter is also caused by two reasons. The
first one is the decrease of the average ionic radius at A-site.
The second one is the decrease of JT distortions in $Mn^{3+}O_6$
octahedra ions due to the increase of Ca doping. $\omega_3$ is
pronounced in LCMO1/STO samples. In LCMO1 films, this pronounced
peak indicates a higher oxygen deficit. As the Ca content
increases to 0.5, it becomes weak. The decrease in the intensity
of this mode does not suggest a decrease of oxygen deficit. It can
be explained by increasing the volume of domains with short-range
charge/orbital-ordered. Therefore, with increasing Ca
concentration, the Raman spectra change differently in the films
on different substrate.

\section{Conclusions}
Raman spectra of LCMO films grown on LAO(001), STO(001), and
MgO(001) substrates were studied at different temperatures.
According to the XRD pattern, there are a compressive strain for
the films on LAO, a tensile strain for the films on STO and a
slight strain for the films on MgO. The variations of Raman
spectra with temperature, doping level and strain are studied. The
effect of temperature and strain are hardly distinguished because
the change of temperature is usually accompanied by a change of
strain. In the films with Ca concentration of 0.33, there is a
PM-FM phase transition as temperature decreases. As the samples
enter FM phase, the FWHM of $\omega_1$ sharpens in LCMO1/LAO. The
intensities of this tilted mode decreases dramatically in LCMO1 on
STO and MgO. These changes of $\omega_1$ are correlated with the
spin-lattice coupling in spin-ordered FM state. For the high
frequency $\omega_4$ mode, an evident increase in their
intensities occurs as temperature decreases. This seems to be a
strain effect and is also observed in LCMO2 films. In the films
with Ca concentration of 0.5, Raman spectra change obviously with
decreasing temperature. The changes indicate that the Raman
spectra of films are strongly correlated with the disorder-order
transition. The increase in the intensity of $\omega_3$ is induced
by the increase of oxygen defects in film on STO due to the large
tensile strain in $La_{0.67}Ca_{0.33}MnO_3$ films. A unsigned mode
appears at high frequency band, which depends on the strain
strongly. As Ca concentration increases, the films on different
substrates present different behaviors.

\begin{acknowledgments}
This work is supported by the Nature Science Foundation of China ,
by the Ministry of Science and Technology of China (Grant No.
NKBRSF-G19990646) and by the Knowledge Innovation Project of
Chinese Academy of Sciences for XHC, and in part by the State of
Texas through the Texas Center for Superconductivity and Advanced
Materials at the University of Houston for CLC.
\end{acknowledgments}

\begin{table*}
\caption{\label{tab:table1}The out-of-plane lattice parameters of
films and substrates estimated from XRD pattern data and
references\cite{Wang1,Schmahl,Mathur,Abrashev}.}
\begin{ruledtabular}
\begin{tabular}{ccccc}
 &\multicolumn{2}{c}{Substrates}&\multicolumn{2}{c}{Films}\\
 Substrate&Exp (\AA)&Ref (\AA)&
$La_{0.67}Ca_{0.33}MnO_3$ (3.86 \AA)(\AA) &$La_{0.5}Ca_{0.5}MnO_3$
(3.84 \AA)(\AA)\\ \hline
 $LaAlO_3$&3.7916&3.79&3.8822&3.8994 \\
 $SrTiO_3$&3.9026&3.905&3.8394&3.8036 \\
 $MgO$&4.2251&4.211&3.8666&3.8466 \\
 \end{tabular}
\end{ruledtabular}
\end{table*}
\noindent

\newpage

{\bf FIGURE CAPTIONS}\\

\noindent Color online: Figure 1: X-ray diffraction patterns for
$La_{1-x}Ca_xMnO_3$ films grown on three substrates with x = 0.33 and 0.5, respectively.\\

\noindent Color online: Figure 2: Raman spectra of
$La_{0.67}Ca_{0.33}MnO_3$ films on different substrates at the
temperatures of 293, 260, 240, 220, 200, and 100 K. In Fig. 2a, the strong feature at about
486 $cm^{-1}$ is the LAO substrate peak identified by asterisks. Insets in Fig. 2 show the
temperature dependence of magnetic moment for films on LAO, STO and MgO, respectively.\\

\noindent Color online: Figure 3: Raman spectra of
$La_{0.5}Ca_{0.5}MnO_3$ films on different substrates at the
temperatures of 293, 250, 200, 180, 160, 140, 100, and 83 K.
In Fig. 3a, the strong feature at about 486 $cm^{-1}$ is the LAO substrate
peak identified by asterisks.\\

\noindent Color online: Figure 4: Raman spectra of as-grown and
annealed LCMO1 films on STO and MgO, and LCMO2 film on LAO,
respectively.

\noindent Color online: Figure 5: Temperature dependence of Raman shift of $\omega_2$, $\omega_3$ and $\omega_4$ for LCMO1 films.\\

\noindent Color online: Figure 6: Temperature dependence of Raman shift of $\omega_2$ and $\omega_4$ for LCMO2 films.\\

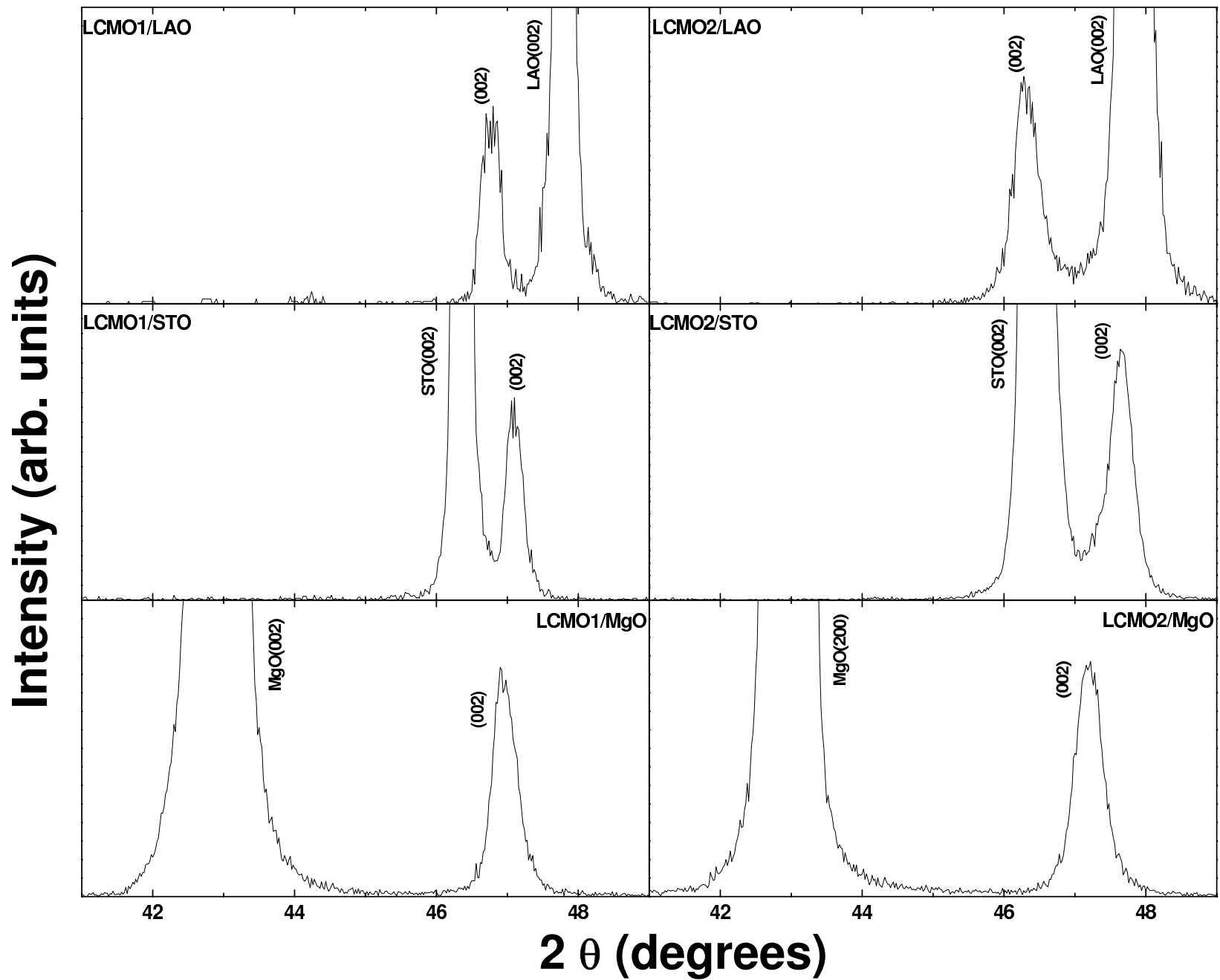
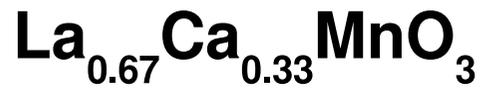
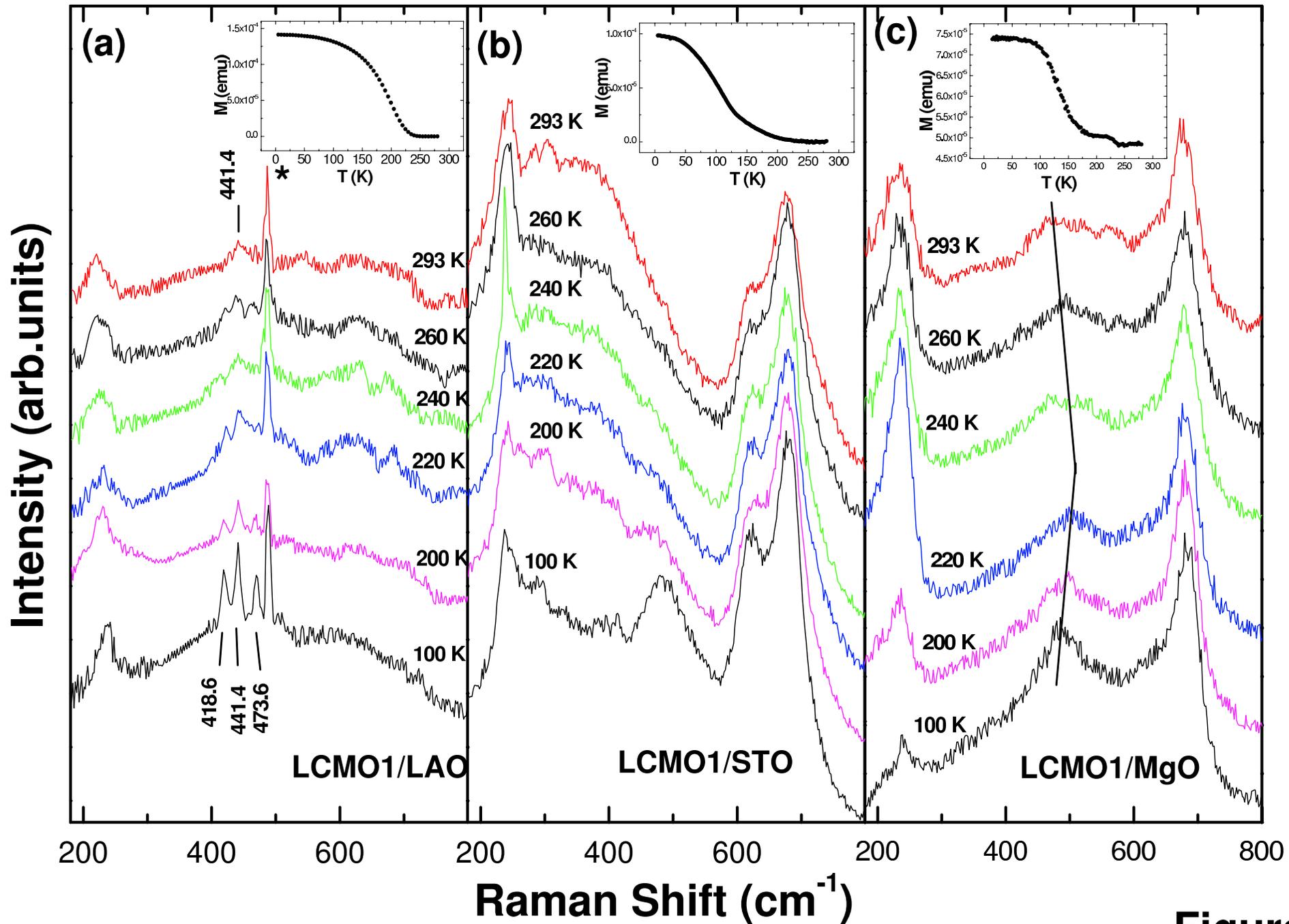
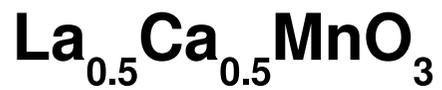
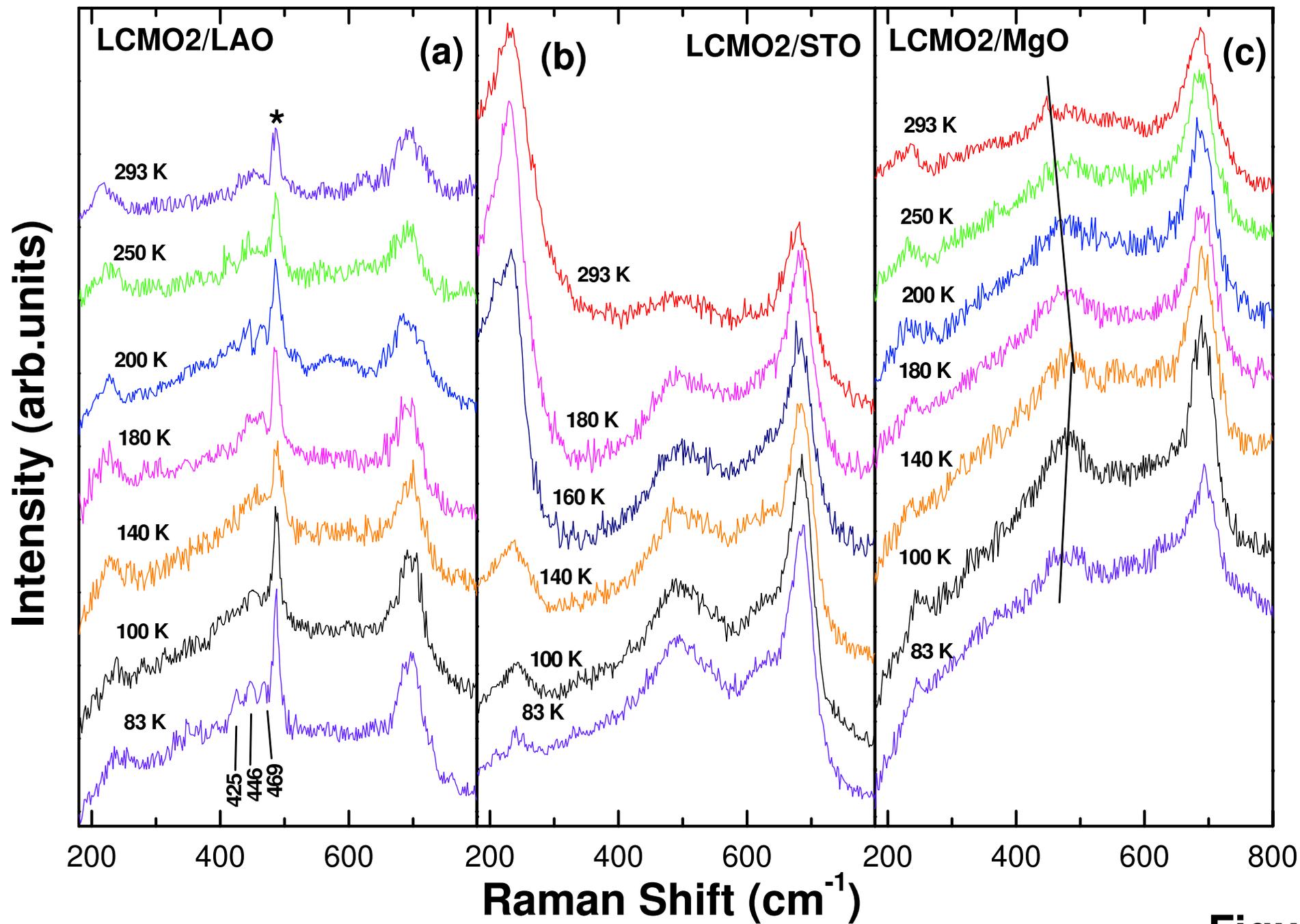
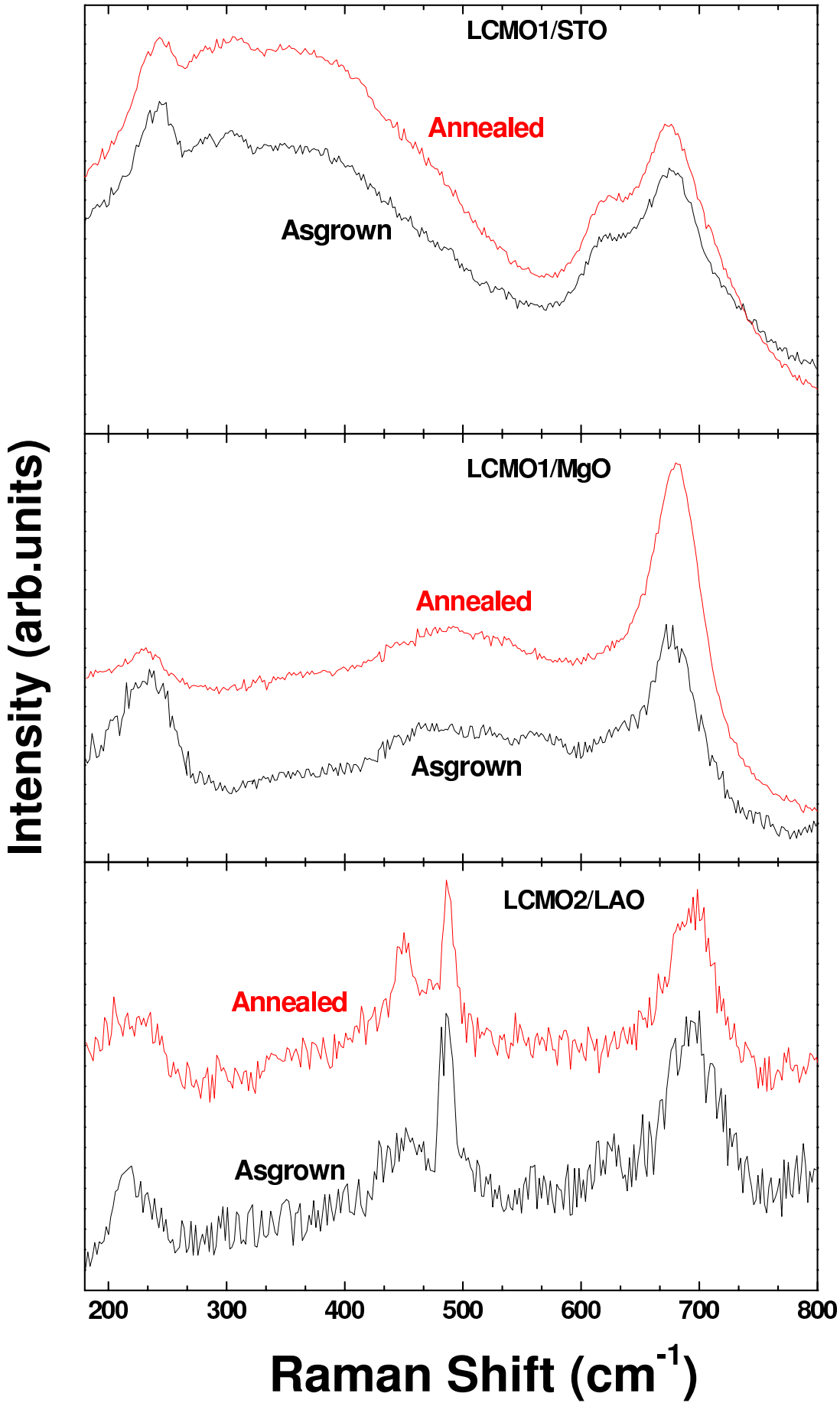
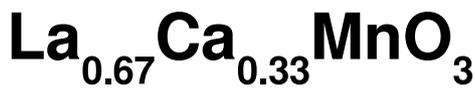
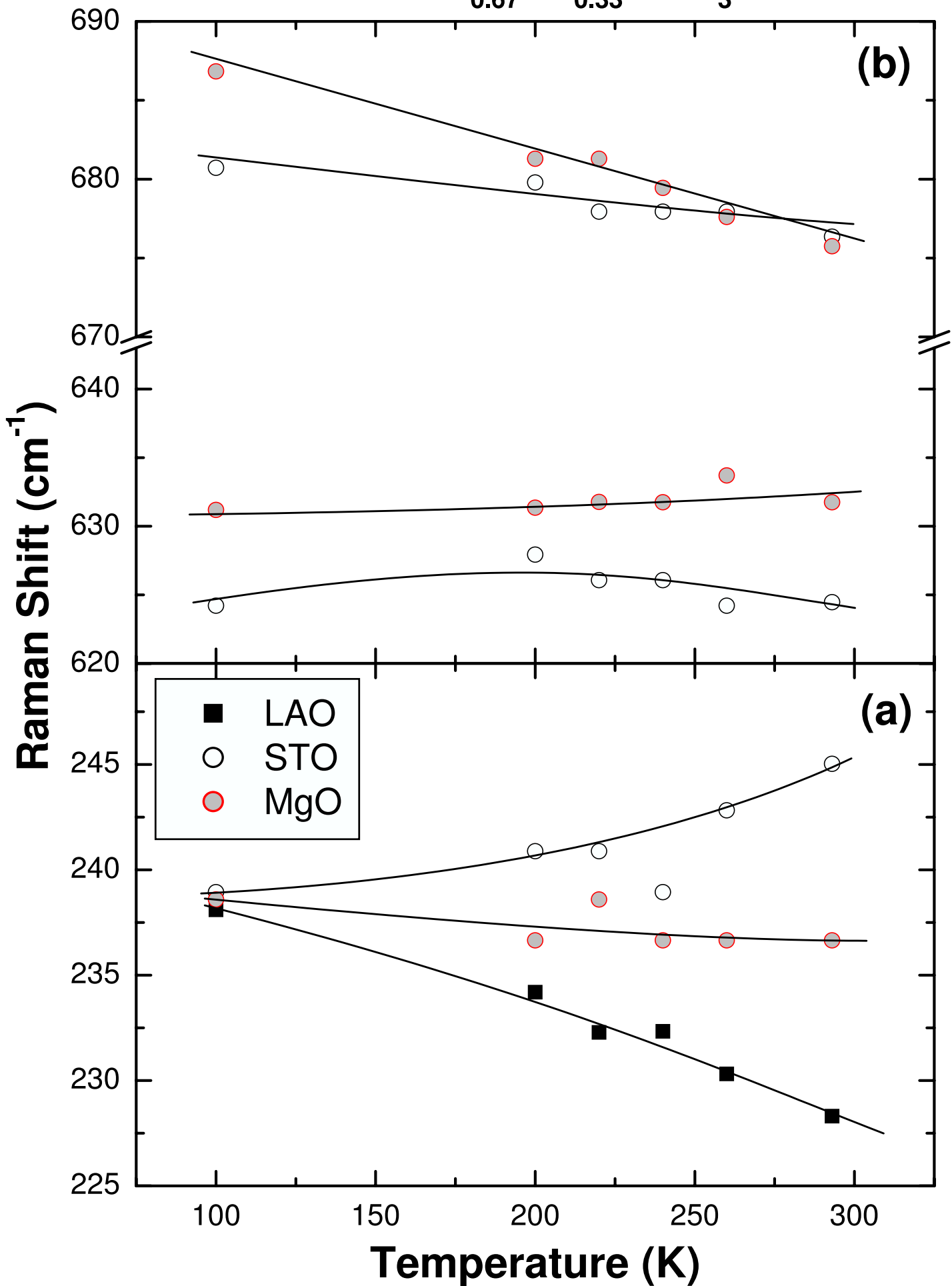
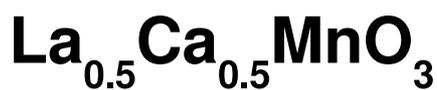
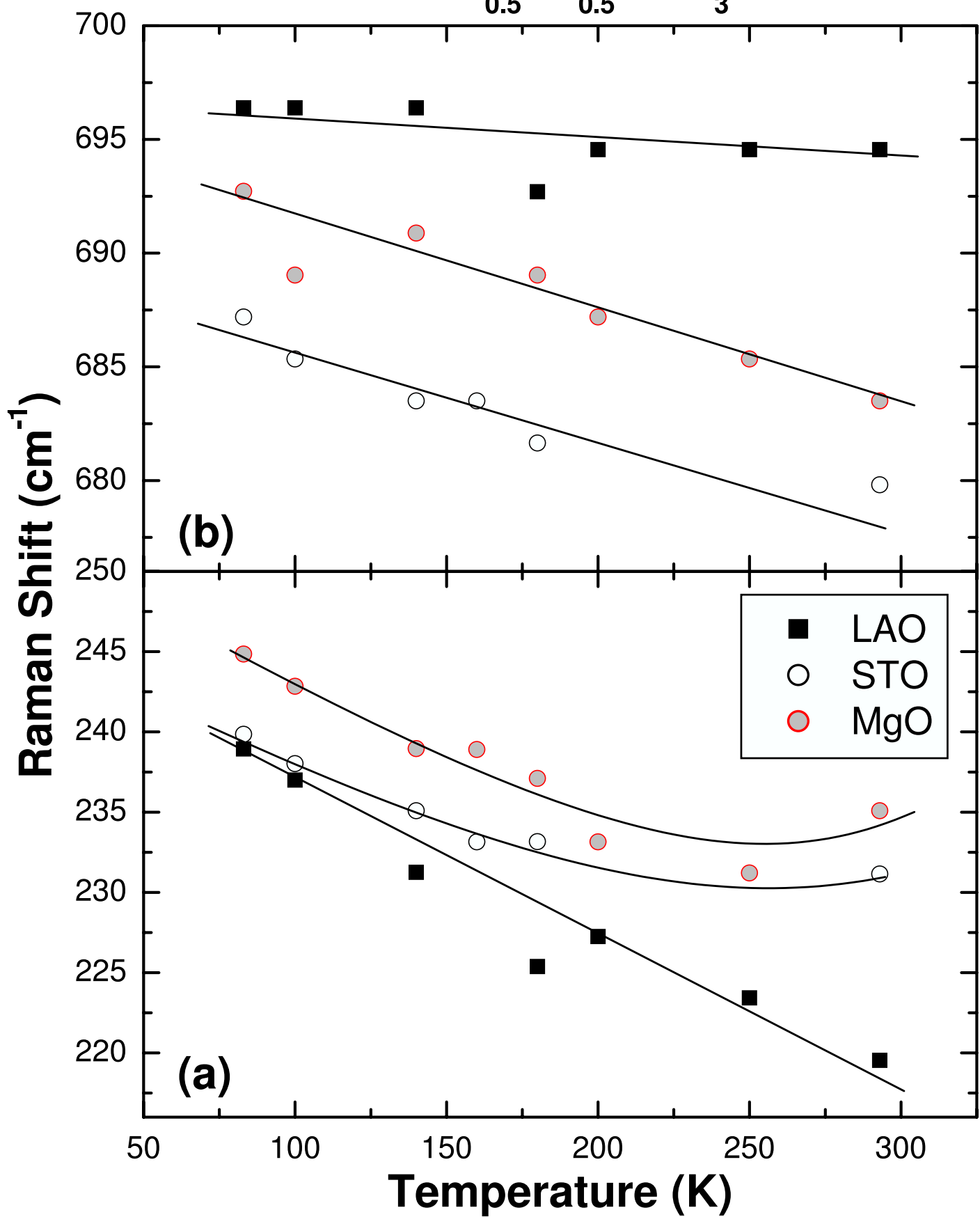
\end{document}